\documentclass[10pt,letterpaper]{article}
\usepackage{opex3}
\usepackage{cite}
\newcommand{\ket}[1]{|#1\rangle}             		% Ket Dirac's notation %
\newcommand{\qo}[1]{``#1''}                     		% quotation %
\newcommand{\op}[1]{\widehat{#1}}         		% operator %
\begin{document}
%
%%%%%%%%%%%%%%%%%% title page information %%%%%%%%%%%%%%%%%%
\title{Generation and dynamics of optical beams with polarization singularities}

\author{Filippo Cardano,$^{1}$ Ebrahim Karimi,$^{1,\ast}$ Lorenzo Marrucci,$^{1,2}$ Corrado de Lisio,$^{1,2}$ and Enrico Santamato$^{1,3}$}
\address{$1$ Dipartimento di Fisica, Universit\`{a} di Napoli ``Federico II'', Complesso di Monte S. Angelo, 80126 Napoli, Italy\\
$2$ CNR-SPIN, Complesso Universitario di Monte S. Angelo, 80126 Napoli, Italy\\
$3$ CNISM-Consorzio Nazionale Interuniversitario per le Scienze Fisiche della Materia, Napoli, Italy}
\email{ekarimi@uottawa.ca}


\begin{abstract}
We present a convenient method to generate vector beams of light having polarization singularities on their axis, via partial spin-to-orbital angular momentum conversion in a suitably patterned liquid crystal cell. The resulting polarization patterns exhibit a C-point on the beam axis and an L-line loop around it, and may have different geometrical structures such as \qo{lemon}, \qo{star}, and \qo{spiral}. Our generation method allows us to control the radius of L-line loop around the central C-point. Moreover, we investigate the free-air propagation of these fields across a Rayleigh range.
\end{abstract}

 % REPLACE WITH CORRECT OCIS CODES FOR YOUR ARTICLE

%%%%%%%%%%%%%%%%%%%%%%% References %%%%%%%%%%%%%%%%%%%%%%%%%

%%%%%%%%%%%%%%%%%%%%%%%%%%  body  %%%%%%%%%%%%%%%%%%%%%%%%%%
\section{Introduction}
It has been widely studied and directly observed that both longitudinal and transverse waves may have structures rich of singular points \cite{nye:74,dennis:09}. Optical light beams with singular scalar and vectorial features received a particular attention \cite{dennis:09}. In the scalar case, these singularities in the beam transverse plane are known as \qo{wave dislocations} or \qo{optical vortices} \cite{nye:74,soskin:01}. They correspond to points where the optical field amplitude vanishes and the optical phase becomes undefined. Near the singular point, the optical phase changes continuously with a specific \textit{topological charge}, given by the phase change in units of $2\pi$ along a closed path surrounding the singularity \cite{soskin:01}. Beams with optical vortices carry orbital angular momentum (OAM), which is simply proportional to the topological charge when the vortex is located on the beam axis \cite{sonja:08}.

\begin{figure}[t]
    \centering
        \includegraphics[width=10cm]{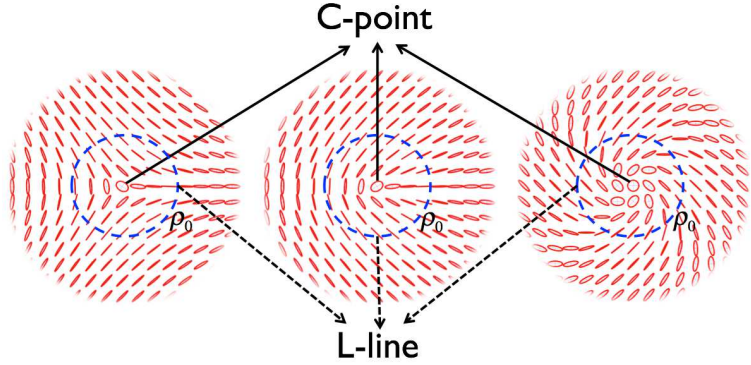}
    \caption{Polarization patterns of three different polarization singular beams: \qo{star} (index $\eta=-1/2$), \qo{lemon} ($\eta=1/2$) and \qo{spiral} ($\eta=1$). The central point at the origin indicates the C-point, while dashed blue lines show the L-line circles with radius $\rho_0$. The handednesses of the polarization ellipses in the outer region, $\rho>\rho_0$, and in the inner one, $\rho<\rho_0$, are opposite.}
    \label{fig:polre}
\end{figure}

When also the vectorial nature of the optical field is taken into account, the variety of possible singular points in the transverse plane is increased \cite{nye:83a,nye:83b,dennis:02,freund:01,angelsky:02,berry:04,soskin:03}. In the paraxial approximation, the optical field is purely transverse, i.e. orthogonal to the propagation direction. In each point of the transverse plane, the direction of the electric field changes in time tracing an ellipse, whose ellipticity and orientation define the local \qo{\textit{polarization}} state of the field. The polarization can be also parametrized by introducing the three \textit{reduced Stokes parameters}, $s_1$, $s_2$ and $s_3$, or the polar and azimuthal angles on the \textit{Poincar\'e sphere}. The vector-field singularities form points or lines in the transverse plane where one of the polarization parameters is undefined. Examples are the so-called \textit{C-points} and \textit{L-lines}, where the orientation and the handedness of the polarization ellipse are undefined, respectively~\cite{nye:83b,dennis:02}. Optical beams exhibiting vector singularities of this kind have been given various names in the literature, such as for example (full)-Poincar\'e beams \cite{beckley:10,galvez:12}. Here, we simply call them \qo{\textit{polarization-singular beams}} (PSB). The PSB may have different transverse polarization patterns, characterized by a different integer or half-integer topological index $\eta$ as defined by the variation of the polarization ellipse orientation for a closed path around a C-point, such as \qo{star}, \qo{monstar}, \qo{lemon}, \qo{spiral} (with a central \qo{source/sink} singularity), \qo{hyperbolic} (with a central \qo{saddle} singularity), etc. Some examples are shown in Fig.\ \ref{fig:polre}. Optical PSB possessing high values of positive and negative topological charge form so-called \qo{polarization flowers} and \qo{hyperbolic webs}, respectively \cite{freund:01}.

Polarization singularities are typically generated either by interferometric methods (see, e.g., \cite{kurzynowski:12}), which is usually very delicate and unstable, or by propagation through suitable birefringent solid crystals (see, e.g., \cite{fadeyeva:12} and references therein), a method which does not lend itself to a flexible control of the output. In this paper, we study the generation and propagation dynamics of vectorial polarization singular beams generated by \qo{q-plates} \cite{marrucci:06a}, i.e. birefringent liquid crystal cells having a singular pattern of the optical axis. In a recent work, we have reported the generation of singular vortex beams with uniform polarization ellipticity and a central vortex singularity with vanishing intensity, as generated by \qo{tuned} q-plates \cite{cardano:12}. Here, by exploiting \qo{untuned} q-plates, we conveniently generate and control beams having a space-variant polarization ellipticity, with a central C-point singularity, and a lemon, star or spiral ellipse-orientation pattern. Finally, we analyze the varying polarization structures in different longitudinal positions along the beam, so as to investigate the effect of propagation.
\section{Polarization singular beams}
A class of polarization singular beams can be generated by the superposition of two optical fields bearing orthogonal polarization states and different OAM values. Here, we consider in particular a beam described by the following expression:
\begin{eqnarray}\label{eq:flower}
    \ket{\hbox{PSB}}=\cos{\left(\frac{\delta}{2}\right)}\,\hbox{LG}_{0,0}(\rho,\phi,z)\,\ket{L}+e^{i\alpha}\sin{\left(\frac{\delta}{2}\right)}\,\hbox{LG}_{0,m}(\rho,\phi,z)\,\ket{R}
\end{eqnarray}
where $\rho,\phi,z$ are cylindrical coordinates, $\delta$ is a tuning parameter, $\ket{L}$ and $\ket{R}$ stand for left- and right-circular polarizations, respectively, $\alpha$ is a phase, and LG$_{p,m}$ denotes a Laguerre-Gauss mode of radial index $p$ and OAM index $m$, with a given waist location ($z=0$) and radius $w_0$.

%
%\begin{eqnarray}\label{eq:lg}
%	\mbox{LG}_{p,m}(\rho,\phi,\zeta)=\sqrt{\frac{ 2^{|m|+1}p!}{\pi(p+|m|)!}}\,\frac{\rho^{|m|}\,e^{-\frac{\rho^2}{1-i\zeta}}}{(1+\zeta^2)^{\frac{1+|m|}{2}}} L_{p}^{|m|}\left(\frac{2\rho^2}{1+\zeta^2}\right) e^{im\phi+i(2p+|m|+1)\arctan{(\zeta)}},
%\end{eqnarray}
%
The LG modes with $m\neq0$ have a doughnut shape, with vanishing intensity at the center. The PSB resulting from Eq.\ (\ref{eq:flower}) then have a polarization pattern, shown in Fig.\ \ref{fig:polre}, characterized by the following four elements: (i) at the center (that is the beam axis), the polarization is left-circular (C-point), since the last term in Eq.~(\ref{eq:flower}) vanishes here; (ii) as the radius increases, the second term contributes and changes the circular polarizations into left-hand elliptical; (iii) at a certain radius $\rho=\rho_0(\delta)$, where the two waves in Eq.~(\ref{eq:flower}) have equal amplitude, the polarization becomes exactly linear (L-lines); (iv) at larger radii, that is for $\rho>\rho_0$, the second term in Eq.~(\ref{eq:flower}) prevails and the polarization state changes into right-handed elliptical. It is worth noting that, in this case, the polarization pattern has a topological index $\eta=m/2$. The initial angle at azimuthal location $\phi=0$ of the polarization ellipse major axis is fixed by the relative phase $\alpha$ in Eq.~(\ref{eq:flower}). Of course, the $\ket{L}$ and $\ket{R}$ polarizations in Eq.~(\ref{eq:flower}) can also be swapped, leading to an inverted polarization ellipticitiy everywhere.
\begin{figure}[t]
    \centering
        \includegraphics[width=9cm]{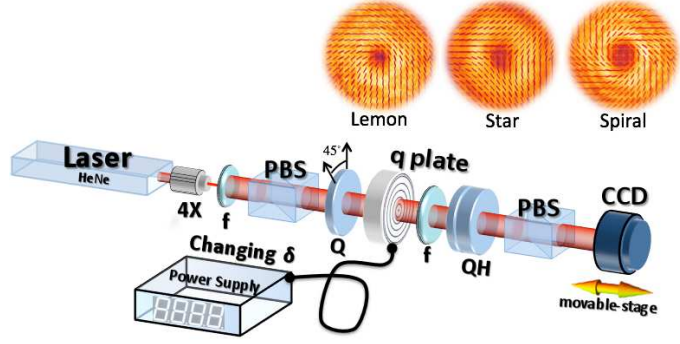}
    \caption{Experimental setup. A He-Ne laser beam is first spatial-mode filtered by focusing via a microscope objective (4X) into a 50~$\mu m$ pinhole and recollimating by a lens (f). Its polarization is then prepared in a left (right) circular state by a polarizing beam-splitter (PBS) and a quarter-wave plate (Q). An untuned q-plate then transforms the beam into a PSB. The PSB polarization pattern was analyzed by a quarter-wave plate, a half-wave-plate (H) and another polarizer, followed by imaging on a CCD camera. The $\delta$ parameter of the PSB was adjusted by electrically tuning the q-plate. In order to study the propagation dynamics of the PSB, the CCD camera was mounted on a translation stage and moved around the focal plane of an output lens. Upper insets show the intensity and reconstructed polarization patterns, in the near field, of the lemon, star, and spiral beams generated by q-plates with $q=1/2,-1/2,1$, respectively. }
    \label{fig:qppatterns}
\end{figure}

We generate PSB by exploiting partial spin-to-orbital angular momentum conversion in q-plates. The optical action of q-plates is described in \cite{marrucci:06a,marrucci:11,karimi:09a,slussarenko:11}. Depending on its birefringent retardation $\delta$, the q-plate converts a varying fraction of the spin angular momentum change suffered by the light passing through the device into OAM with $m=\pm2q$, where $q$ is the q-plate topological charge and the sign is determined by the input circular polarization handedness. The conversion is full when the q-plate optical phase retardation is $\delta=\pi$ (or any odd multiple of $\pi$). For other values of $\delta$, the q-plate is said to be untuned and the spin-to-orbital conversion is only partial. In the untuned q-plate a fraction of the incoming beam passes through unchanged. The field at the exit plane of an untuned q-plate for a Gaussian left-circular polarized input beam is given by the following expression:
\begin{eqnarray}\label{eq:q-plate}
    \op{\mbox{U}}\cdot\ket{L}=\hbox{HyGG}_{-|q| \sqrt{2},q\sqrt{2}}\left(\rho,d/\bar{n}\right)\left[\cos{\left(\frac{\delta}{2}\right)}\ket{L}+e^{i\alpha}\sin{\left(\frac{\delta}{2}\right)}\ket{R}\,e^{2iq\,\phi}\right],
\end{eqnarray}
where $\hbox{HyGG}_{-|q|\sqrt{2},q\sqrt{2}}$ denotes the amplitude profile of a hypergeometric-Gaussian beam \cite{karimi:07}, describing with good approximation the output mode profile of a q-plate of charge $q$, thickness $d$ and average refractive index $\bar{n}$ for a Gaussian input (see Ref.\ \cite{karimi:09b} for more details), and the phase $\alpha$ in the second term depends on the initial angle (on the $x$ axis) of the q-plate optical-axis pattern \cite{marrucci:06a,marrucci:11}. The converted and non-converted terms in Eq.~(\ref{eq:q-plate}), that is in the near field, have a common amplitude profile, which can therefore be factorized. The far-field profiles produced by the converted and non-converted parts of the beam are different, however, because the different OAM values affect the beam propagation. The HyGG term appearing in Eq.~(\ref{eq:q-plate}), after multiplication by the phase factor $\exp{(2iq\phi)}$, can be expanded in a series of LG modes \cite{karimi:07}. Truncating this expansion to the first useful order (which is a good approximation in the far field), Eq.~(\ref{eq:q-plate}) is reduced to Eq.~(\ref{eq:flower}) with $m=2q$. The optical retardation $\delta$ of the q-plate, that is the q-plate tuning, can be adjusted continuously by applying an appropriate voltage \cite{slussarenko:11}. Different polarization topologies can be obtained by using q-plates with different topological charges. In this work, we used $q=-1/2,+1/2$, and $+1$, generating correspondingly equal indices of the polarization patterns.
%%%
\section{Experiment}
\begin{figure}[t]
    \centering
        \includegraphics[width=13cm]{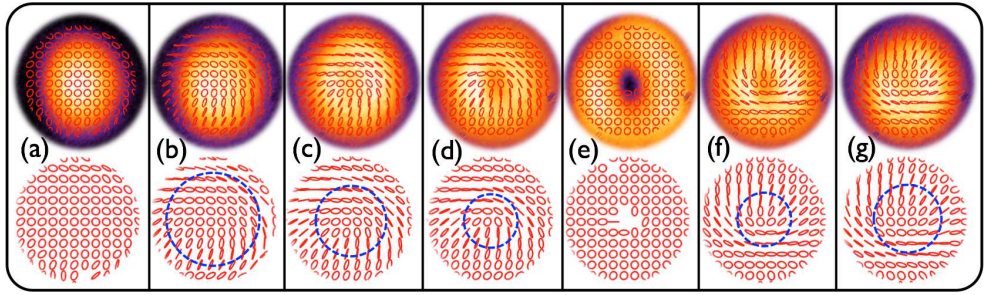}
    \caption{Intensity distribution and reconstructed polarization patterns of beam generated by a q-plate with $q=1/2$ for seven different optical retardations: (a) $\delta=0$ (or $2\pi$), (b) $\delta=\pi/4$, (c) $\delta=\pi/2$, (d) $\delta=3\pi/4$, (e) $\delta=\pi$, (f) $\delta=5\pi/4$, and (g) $\delta=3\pi/2$. The corresponding L-line radii relative to the beam waist $w_0$ are the following: (a) undefined (b) $\rho_0=2.0 w_0$, (c) $\rho_0=1.4 w_0$, (d) $\rho_0=1.1 w_0$, (e) undefined, (f) $\rho_0=1.0 w_0$, and (g) $\rho_0=1.5 w_0$.}
    \label{fig:qpcregion}
\end{figure}

The experiment layout is shown in Fig.\ \ref{fig:qppatterns}. The generated PSB polarization distribution was studied both in the near and far fields generated in the image and focal planes of a lens. The polarization distribution was analyzed by polarization tomography: the local Stokes parameter maps have been calculated from the intensity images taken by a CCD camera (Sony AS-638CL), after projection onto horizontal ($H$), vertical ($V$), anti-diagonal ($A$), diagonal ($D$), $L$ and $R$ polarization states, respectively.  The polarization pattern was thus entirely reconstructed, as shown in form of little ellipses in Fig.\ \ref{fig:qppatterns} and subsequent figures. In order to reduce noise and experimental errors, the average intensity was recorded in a grid of 20$\times$20 pixel square area~\cite{cardano:12}. The missing ellipses in the experimentally reconstructed polarization patterns (e.g., in Fig.~\ref{fig:qpcregion}) correspond to regions with low intensity, so that the polarization reconstruction was too affected by stray light.
The superposition coefficients in Eq.~(\ref{eq:q-plate}) have been adjusted by changing the q-plate optical retardation $\delta$ with the applied voltage. The radius where the L-line forms in the polarization pattern depends on the ratio between the absolute value of the two coefficients in Eq.~(\ref{eq:q-plate}) and, hence, on the voltage applied to the q-plate. In particular, for optical retardation $\delta=0$ and $\delta=\pi$ the beam profile is Gaussian fully left-circular and doughnut-shaped fully right-circular, respectively. In the intermediate case, the pattern is a superposition and a circular L-line appears in the polarization pattern, whose radius $\rho_0$ depends on $\delta$. From Eq.\ (\ref{eq:flower}) one finds $\rho_0= w_0/(\sqrt{2} \tan{\delta/2})$. Figure (\ref{fig:qpcregion}) shows the intensity and reconstructed polarization pattern for different q-plate retardations in steps $\Delta\delta=\pi/4$, and the corresponding measured L-lines radii. As shown in the figure, also the intensity profile changes from a Gaussian (left-circular) to a doughnut (right-circular) shape. In Fig.~\ref{fig:qpcregion} (c) and (g), the optical retardations are $\pi/2$ and $3\pi/2$, respectively; the $\pi$ difference in the relative phases results into a $180^{\circ}$ rotation of the polarization patterns. The topological index of the polarization pattern, however, depends only on the $q$-value of the q-plate, and in the case of Fig.~\ref{fig:qpcregion} we have $\eta=q=1/2$.

\begin{figure*}[t]
    \centering
        \includegraphics[width=13.5cm]{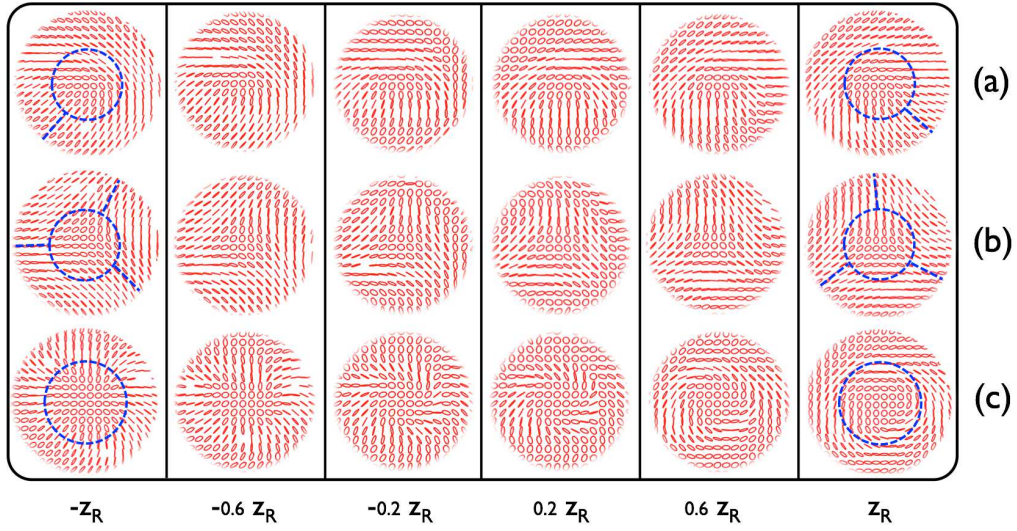}
    \caption{Reconstructed experimental polarization patterns of different PSB beams. (a) $\eta=m/2=+1/2$, (b) $\eta=m/2=-1/2$ and (c) $\eta=m/2=+1$. Patterns have been reconstructed by measuring the maps of reduced Stokes parameters in six different longitudinal planes within the beam Rayleigh range, from $-z_R$ to $+z_R$. The corresponding rotation of polarization patterns for (a) and (b) are $90^{\circ}$ and $88^{\circ}=(30+30+28)^{\circ}$, respectively.}
    \label{fig:free-air}
\end{figure*}

In a second experiment, we studied the dynamics of different PSBs with $\eta=-1/2$, $+1/2$ and $+1$ topologies under free-air propagation for fixed parameter $\delta=\pi/2$. To this purpose, we moved the CCD camera along the propagation axis around the imaging-lens focal plane. We recorded the beam polarization patterns at six different planes in the range $-z_R\le z\le z_R$, where $z_R$ is the lens Rayleigh parameter and $z=0$ corresponds to the beam waist location. The experimental results are shown in Fig.\ \ref{fig:free-air}. As it can be seen, the polarization pattern evolves during propagation. Indeed  LG$_{00}$ and LG$_{0,m}$ modes of Eq. (\ref{eq:flower}) have different $z$ dependences of their Gouy phases; the relative Gouy phase between them is given by $\psi=|m|\arctan{(z/z_0)}$, which in the explored region varies in the range $|m|[-\pi/4,\pi/4]$. In the case $m=\pm 1$ (i.e.\ $\eta=1/2$) the phase evolution leads to a rigid rotation of the whole polarization structure by an angle equal to $\psi$ [as shown in Fig.~\ref{fig:free-air} (a) and (b)], while when $m=2$ (i.e.\ $\eta=1$), this dephasing leads to a more complex pattern dynamics, in which the polarization structure changes from radial to spiral and then to azimuthal (Fig.~\ref{fig:free-air} (c)).
%
%%%%
\section{Conclusion}
We reported a non-interferometric technique to generate beams carrying a polarization singularity on their axis, based on a partial spin-to-orbital angular momentum conversion in a detuned q-plate device. By this approach, we generated beams with C-points on their optical axis, surrounded by a circular L-line whose radius can be controlled by the voltage applied to the q-plate. We demonstrated different patterns such as stars, lemons, and spirals, corresponding to vector topological indices of $-1/2$, $1/2$ and $+1$, respectively. We have also studied the pattern evolution in free-space propagation. Such polarization patterns could find practical application in lithographic optical techniques, as for the examples discussed in Refs.\ \cite{hnatovsky:11,ambrosio:12}.
%%%%
\section*{Acknowledgments}
We thank Sergei Slussarenko for preparing the q-plates. We acknowledge financial support of the Future and Emerging Technologies FET-Open programme, within the $7^{th}$ Framework Programme for Research of the European Commission, under grant number 255914 - PHORBITECH.

\begin{thebibliography}{99}

\bibitem{nye:74} % original starting paper on singular beam
J. F. Nye and M. V. Berry, \qo{Dislocations in Wave Trains,} Proc. R. Soc. Lond. A {\bf 336,} 165--190 (1974).

\bibitem{dennis:09} % review on vortices and singularities
 M. R. Dennis, K. O'Holleran, and M. J. Padgett, \qo{Optical Vortices and Polarization Singularities,} Prog. Opt. {\bf 53,} 293--363 (2009).

\bibitem{soskin:01} % singular optics review
M. S. Soskin and M. V. Vasnetsov (Ed. E. Wolf), \qo{Singular optics,} Prog. Opt. {\bf 42,} 219--276 (2001).

\bibitem{sonja:08} % OAM review
S. Franke-Arnold, L. Allen, and M. J. Padgett, \qo{Advances in optical angular momentum,} Laser Photonics Rev. {\bf 2,} 299--313 (2008).

\bibitem{nye:83a} % disclinations in e.m. waves
J. F. Nye,  \qo{Polarization effects in the diffraction of electromagnetic waves: the role of disclinations,} Proc. Roy. Soc. Lond. A  {\bf 387,} 105--132 (1983).

\bibitem{nye:83b} % c-points in e.m. waves
J. F. Nye,  \qo{Lines of circular polarization in electromagnetic wave fields,} Proc. Roy. Soc. Lond. A  {\bf 389,} 279--290 (1983).

\bibitem{dennis:02} % c-points morphology and statistic in random fields
M. R. Dennis,  \qo{Polarization singularities in paraxial vector fields: morphology and statistics,} \oc  {\bf 213,} 201--221 (2002).

\bibitem{freund:01} % polarization flowers
I. Freund, \qo{Polarization flowers,} \oc {\bf 199,} 47--63 (2001).

\bibitem{angelsky:02} % polarization singularities and component vortices
O. Angelsky, A. Mokhun, I. Mokhun, and M. Soskin, \qo{The relationship between topological characteristics of component vortices and polarization singularities,} \oc  {\bf 207,} 57--65 (2002).

\bibitem{berry:04} % polarization singularity in nature
M. V. Berry, M. R. Dennis, and R. L. Lee Jr., \qo{Polarization singularities in the clear sky,} New J.\ Phys.\ {\bf 6,} 162 (2004).

\bibitem{soskin:03} % dynamic of singularity
M. S. Soskin, V. Denisenko, and I. Freund, \qo{Optical polarization singularities and elliptic stationary points,} \ol {\bf 28,} 1475--1477 (2003).

\bibitem{beckley:10} % Poincare beam
 A. M. Beckley, T. G. Brown, and M. A. Alonso, \qo{Full Poincare beams,} \opex {\bf 18,} 10777-10785 (2010).

\bibitem{galvez:12} % Poincare beam
E. J. Galvez, S. Khadka, W. H. Schubert, and S. Nomoto, \qo{Poincar{\'e}-beam patterns produced by nonseparable superpositions of Laguerre-Gauss and polarization modes of light,} \ao {\bf 51,} 2925--2934 (2012).

\bibitem{kurzynowski:12} % interference of three waves
P. Kurzynowski, W. A. Wo\'{z}niak, M. Zdunek, and M. Borwi\'{n}ska, \qo{Singularities of interference of three waves with different polarization states,} \opex {\bf 20,} 26755--26765 (2012).

\bibitem{fadeyeva:12} % 
T. A. Fadeyeva, C. N. Alexeyev, P. M. Anischenko, and A. V. Volyar, \qo{Engineering of the space-variant linear polarization of vortex-beams in biaxially induced crystals,} \ao {\bf 51,} C224–-C230 (2012).

\bibitem{marrucci:06a} % q-plate
L. Marrucci, C. Manzo, and D. Paparo, \qo{Optical Spin-to-Orbital Angular Momentum Conversion in Inhomogeneous Anisotropic Media,} \prl {\bf 96,} 163905 (2006).

\bibitem{cardano:12} % q-plate output beam pattern
 F. Cardano, E. Karimi, S. Slussarenko, L. Marrucci, C. de Lisio, and E. Santamato, \qo{Polarization pattern of vector vortex beams generated by q-plates with different topological charges,} \ao  {\bf 51,} C1--C6 (2012).

%\bibitem{siegman:86}
%A.~E.~Siegman, {\it Lasers} (University Science Books, 1986).

\bibitem{marrucci:11} % q-plate review
L. Marrucci, E. Karimi, S. Slussarenko, B. Piccirillo, E. Santamato, E. Nagali, and F. Sciarrino, \qo{Spin-to-orbital conversion of the angular momentum of light and its classical and quantum applications,} J. Opt.\ {\bf 13,} 064001 (2011).

\bibitem{karimi:09a} % q-plate tuning
E. Karimi, B. Piccirillo, E. Nagali, L. Marrucci, and E. Santamato, \qo{Efficient generation and sorting of orbital angular momentum eigenmodes of light by thermally tuned q-plates,} \apl {\bf 94,} 231124 (2009).

\bibitem{slussarenko:11} % q-plate tuning
S. Slussarenko, A. Murauski, T. Du, V. Chigrinov, L. Marrucci, and E. Santamato, \qo{Tunable liquid crystal q-plates with arbitrary topological charge,} \opex {\bf 19,} 4085-4090 (2011).

\bibitem{karimi:07} % HyGG beam
E. Karimi, G. Zito, B. Piccirillo, L. Marrucci, and E. Santamato, \qo{Hypergeometric-Gaussian modes,} \ol {\bf 32,} 3053--3055 (2007).

\bibitem{karimi:09b} % output beam from q-plate
E. Karimi, B. Piccirillo, L. Marrucci, and E. Santamato, \qo{Light propagation in a birefringent plate with topological charge,} \ol {\bf 34,} 1225--1227 (2009).

%\bibitem{freund:10} % detect the polarization pattern
%I. Freund, \qo{Multitwist optical M{\"o}bius strips,} \ol {\bf 35,} 148--150 (2010)

\bibitem{hnatovsky:11} % detect the polarization pattern
C. Hnatovsky, V. Shvedov, W. Krolikowski, and A. Rode, \qo{Revealing Local Field Structure of Focused Ultrashort Pulses,} \prl {\bf 106,} 123901 (2011).

\bibitem{ambrosio:12} % detect the polarization pattern
A. Ambrosio, L. Marrucci, F. Borbone, A. Roviello, and P. Maddalena, \qo{Light-induced spiral mass transport in azo-polymer films under vortex-beam illumination,} Nat.\ Commun.\ {\bf 3,} 989 (2012).

%\bibitem{toyoda:12} % detect the polarization pattern
%K. Toyoda, K. Miyamoto, N. Aoki, R. Morita, and T. Omatsu, \qo{Using Optical Vortex To Control the Chirality of Twisted Metal Nanostructures,} Nanolett.\ {\bf 12}, 3645--3649 (2012).
\end{thebibliography}
\end{document}